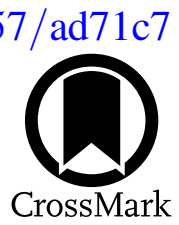

# Exploring the Small-scale Magnetic Fields in the Atmosphere of HD 49385 by Asteroseismic Analysis

Yuetong Wang[1,2,3], Yaguang Li[4], Yan Li[1,2,3,5], Guifang Lin[2,3], and Tao Wu[2,3,5]
[1] Yunnan Observatories, Chinese Academy of Sciences, Kunming 650216, People's Republic of China
[2] University of Chinese Academy of Sciences, Beijing 100049, People's Republic of China; ly@ynao.ac.cn
[3] International Center of Supernovae at the Yunnan Key Laboratory, Kunming 650216, People's Republic of China
[4] Center for Astronomical Mega-Science, Chinese Academy of Sciences, Beijing 100012, People's Republic of China
[5] Institute for Astronomy, University of Hawai'i, 2680 Woodlawn Drive, Honolulu, HI 96822, USA

## Abstract

Recent asteroseismic studies have shown convincing evidence that magnetic fields may exist in the interior of some pulsating red giants. Inspired by this breakthrough, we explored the effect of small-scale magnetic fields on the $p$-mode oscillations in an evolved star, HD 49385. We incorporate a modified Eddington $T$–$\tau$ equation that phenomenologically mimics the effect of the magnetic fields in the atmosphere of HD 49385, and calculate the frequencies of $p$-modes with $l = 0$, 1, and 2. By comparing the calculated frequencies with the observed ones, we select two best-fit models with either GS98 or A09 chemical composition. Our best-fit models not only fit satisfactorily the observed frequencies but also well reproduce some spectroscopically observed stellar parameters such as effective temperature and log $g$. Based on the two best-fit models, we have estimated that the small-scale magnetic fields possess a strength of approximately 80 G and spread concentratively at approximately a height of 1850 km in the atmosphere. By selecting the best-fit models with a special requirement on the avoided-crossing mode, we have confirmed that the frequency of the avoided-crossing mode is tightly related to the helium core of the star, and determined the size of the helium core as 0.117 $M_\odot$ in mass and 0.078 $R_\odot$ in radius. Based on the improvements of the previous two sides, we can accurately determine the mass of HD 49385 to be $1.25 \pm 0.02\ M_\odot$ with an age of 4.1 Gyr for GS98 composition and 4.5 Gyr for A09 composition.

## 1. Introduction

With the use of space telescopes such as CoRoT (A. Baglin et al. 2006) and Kepler (D. G. Koch et al. 2010), highly precise and long-lasting photometric observations have become increasingly available, providing stronger support for the development of asteroseismology. By fitting the eigenfrequencies of the stellar models with those of the corresponding ones of the observed star, some basic physical parameters of the star such as mass, radius, luminosity, and age can be given accurately (D. Stello et al. 2009; T. Kallinger et al. 2010; S. Basu et al. 2011; Y. Wang et al. 2023). By analyzing the frequency splitting of the observed oscillation modes, the rotation and magnetic field of the star can be measured (Y. Li et al. 2021; G. Li et al. 2022, 2023; S. Deheuvels et al. 2023). By detecting the deviation of the observed oscillation mode from the equal frequency-spacing behavior, the hierarchical structure of elements in the interior of stars can be explored (S. Deheuvels et al. 2010; S. Deheuvels & E. Michel 2011). By matching averaged 3D hydrodynamic envelope models to standard 1D equilibrium models over a grid on the $T_{\rm eff}$–log $g$ plane, T. Sonoi et al. (2019) investigated the effect of near-surface convection on the eigenfrequencies of solar-like pulsating stars and proposed a new formula for the so-called surface effect.

Magnetic fields exist extensively in the interior of stars and show a significant effect in the whole lifetime of stars. For the Sun and solar-like stars, the stellar dynamo operates in their convective envelope and leads to observational phenomena such as stellar flares and coronal mass ejections. The ejected plasma will travel along the magnetic field lines and create a torque on the star, which regulates the spin-down of the star (A. Skumanich 1972; S. A. Barnes 2003, 2010). For more massive stars, convection occurs mainly in the central region of the star, which can act vigorously as a magnetic dynamo. According to numerical simulations, this magnetic field can be very strong (more than $10^6$ G; K. C. Augustson et al. 2016). This internal magnetic field cannot be observed directly. However, its interaction with internal gravity waves, which are excited at the surface of the convective core and propagating outwards into the stellar envelope, will be an efficient mechanism for the transport of angular momentum and the mixing of matter (M. Cantiello et al. 2014; J. Fuller et al. 2015; J. C. Wheeler et al. 2015).

Recently, based on an analysis of the asymmetry of dipole mixed-mode multiplets of three pulsating red giants, G. Li et al. (2022) found convincing evidence of 30–100 kG magnetic fields in the vicinity of the hydrogen-burning shell of these stars. At the same time, they also suggested from the measurements that KIC 8684542 has a dipolar magnetic field aligned with the rotation axis, and KIC 11515377 also has a dipolar magnetic field but aligned with the equator. By analyzing the deviation of dipole mixed-mode period sequence from the equally periodic-spacing feature of 11 pulsating red

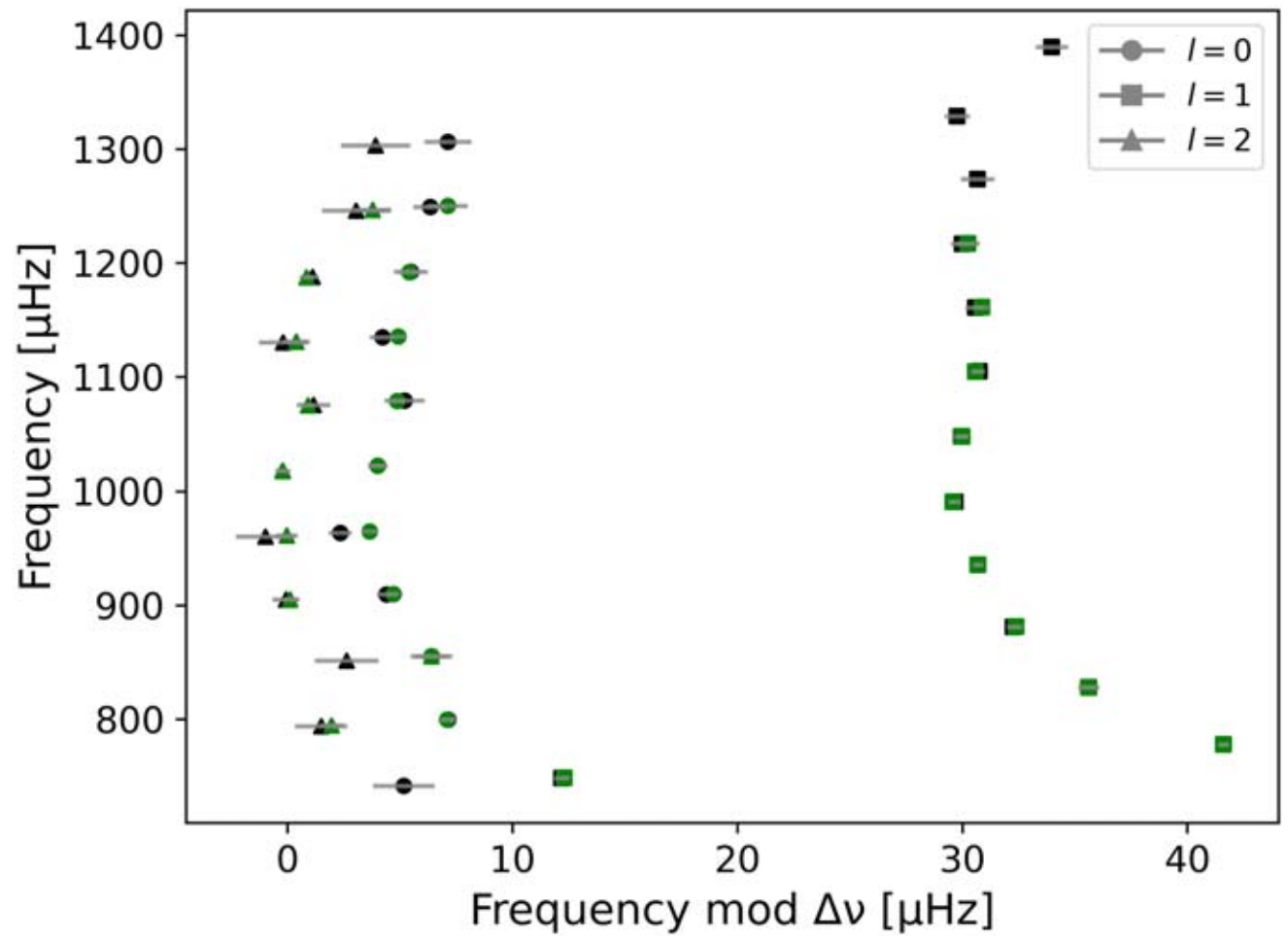

**Figure 1.** Differences between frequencies extracted in this work (black symbols) and in S. Deheuvels et al. (2010; green symbols) for HD 49385. The gray horizontal short lines represent the error bar of the observed frequencies.

giants, S. Deheuvels et al. (2023) found evidence of strong magnetic fields in their core. On the other hand, small-scale magnetic fields are thought to be present in the solar photosphere as well as in the atmosphere of solar-like stars (J. O. Stenflo 1987, 2013). Y. Li et al. (2021) incorporated these small-scale magnetic fields into the solar atmosphere model and found that the agreement of the calculated and observed frequencies of the solar *p*-mode oscillations is greatly improved. Moreover, the derived intensity (90 G) and location (630 km above the photosphere) of the magnetic fields are in good agreement with observations (B. W. Lites et al. 1996, 2008; J. Trujillo Bueno et al. 2004) and numerical simulations (O. Steiner et al. 2008; M. Rempel 2014).

HD 49385 is the first G0-type solar-like star observed by the space telescope CoRoT. S. Deheuvels et al. (2010) analyzed its 137 days of data and showed that HD 49385 is a post-main-sequence star whose oscillations belong to solar-like ones. Additionally, HD 49385 exhibits a very interesting phenomenon, namely the avoided crossing. The avoided crossing refers to the process where two interacting oscillation modes exchange their properties as the eigenfrequency of a *g*-mode approaches the eigenfrequency of a *p*-mode (M. Aizenman et al. 1977). A molecular weight gradient region forms as a result of the reduction of the convection in the stellar core during the main-sequence evolution. Therefore, the avoided crossing serves as an excellent indicator of the presence and nature of the molecular weight gradient region. S. Deheuvels & E. Michel (2011) conducted an asteroseismological work on HD 49385 by constructing a series of theoretical stellar models and then matching the frequencies of their oscillation modes with the frequencies of the observed oscillation modes, especially the fit of the two modes affected by the avoided crossing, to determine the acceptability of the final theoretical model. However, all of the best-fit theoretical models currently available are unable to match all the asteroseismic signals observed in HD 49385, even after applying empirical formula corrections (e.g., W. H. Ball & L. Gizon 2014, and references therein) for the surface effect. In the present work, we introduce a modified Eddington $T$–$\tau$ relation to phenomenologically mimic the effect of the magnetic fields in the atmosphere of HD 49385, and perform asteroseismic analysis to investigate how these magnetic fields would affect the propagation and reflection of stellar oscillations and the effect on their frequencies. This approach can serve as a new method to probe the small-scale magnetic fields in the photosphere of remote solar-like stars. Additionally, we want to precisely determine the mass of the helium core for HD 49385, which is a tight constraint on the convective core overshooting in its main-sequence stage.

In Section 2, we present our method to extract the frequencies of the oscillation modes for HD 49385. In Section 3, we introduce our asteroseismic models, including the input physics on which the stellar models are based, the small-scale magnetic fields that are introduced into the stellar atmosphere, the boundary condition that is adopted in the calculations of the frequencies of the oscillation modes, and finally the fitting method to choose the best-fit model. Then, we show our fitting results for HD 49385 in Section 4 and summarize the conclusions in Section 5.

## 2. Extraction of Oscillation Frequencies for HD 49385

We extracted the oscillation frequencies of HD 49385 following the procedures described in Y. Li et al. (2020). We treated the power spectrum as a combination of granulation background and individual oscillation modes. The identification of mode $n$- and $\ell$- degrees was straightforward due to the high signal-to-noise ratio of the power spectrum. The profile of each oscillation mode ($n$, $\ell$) was modeled as a Lorentzian function, based on the assumption that the modes are well resolved. This assumption is valid given the observing length $T_{\rm obs} = 137$ days, which significantly exceeds the typical mode lifetimes in subgiants, usually a few days. The frequency, full width at half maximum, and amplitude of each Lorentzian profile were treated as free parameters with uniform priors. These mode parameters were jointly optimized using a Bayesian framework with Markov Chain Monte Carlo sampling algorithms (see also G. R. Davies et al. 2016; E. Corsaro et al. 2020).

We compare our results of frequency extraction with those of S. Deheuvels et al. (2010) in Figure 1. It can be noticed that the agreement between the two results is mostly in 1$\sigma$, except for one of $l = 0$ mode ($963.84 \pm 0.51\,\mu$Hz in our work, while $965.16 \pm 0.36\,\mu$Hz in S. Deheuvels et al. 2010) and another one of $l = 2$ mode ($851.52 \pm 1.42\,\mu$Hz in our work, but $855.29 \pm 0.93\,\mu$Hz in S. Deheuvels et al. 2010). The signals of these two modes are quite weak in the original power spectrum, and their relatively large error bars make them have smaller significance in the model fitting.

## 3. Asteroseismic Models

### 3.1. Input Physics

Our stellar models were created using the Modules for Experiments in Stellar Astrophysics (MESA, version 11554). This software was initially released by B. Paxton et al. (2011), with subsequent versions released since then (B. Paxton et al. 2013, 2015, 2018, 2019). Among the input parameters for our modeling, we incorporated the equation of state (EOS) published by F. J. Rogers & A. Nayfonov (2002), as well as the opacity tables published by C. A. Iglesias & F. J. Rogers (1996) for the high-temperature region and by J. W. Ferguson et al. (2005) for the low-temperature region. Two chemical compositions, namely GS98 (N. Grevesse & A. J. Sauval 1998) and A09 (M. Asplund et al. 2009), were employed to construct the corresponding stellar models. Additionally, we modeled the stellar atmosphere using the Eddington gray atmosphere model, with the inclusion of specific magnetic fields described in the next subsection. The standard mixing-length theory was used to treat the convection, and the mixing-length parameter was used as a free parameter. The overshooting mixing was not included in our stellar models, as well as the mixing effect due to element diffusion.

To calculate the stellar evolution models and pulsation frequencies, we used the "adipls" module of the MESA program. This module is an adiabatic oscillation program described by J. Christensen-Dalsgaard (2008).

### 3.2. Adding Magnetic Fields

In our work, we employ the Eddington gray atmosphere model as the atmospheric boundary condition (C. Aerts et al. 2010). The $T$–$\tau$ relation in the atmosphere integration can be written in a unified form:

$$T^4 = \frac{3}{4} T_{\rm eff}^4 [\tau + q(\tau)], \tag{1}$$

where $T$ represents temperature, $T_{\rm eff}$ represents the stellar effective temperature, $\tau$ represents the optical depth from the stellar surface, and $q(\tau)$ is a function similar to Hopf-like function. From this temperature distribution formula, we can see that temperature decreases as the optical depth decreases, reaching its minimum at the surface of the star. The decrease in temperature corresponds to reductions in pressure and density throughout the atmosphere.

In MESA calculations, once the temperature is known from the $T$–$\tau$ relation, pressure can be calculated from the hydrostatic equilibrium equation. The calculated pressure includes both gas pressure and radiation pressure. The gas pressure is obtained from the EOS, while the radiation pressure depends solely on the fourth power of temperature. Therefore, we plan to adjust the temperature, which simulates the effect of magnetic fields in the upper atmosphere.

Y. Li et al. (2021) introduced an additional term in the Hopf-like function in our traditional $T$–$\tau$ relation:

$$q(\tau) = \frac{2}{3} + a \exp(-b\sqrt{\tau}). \tag{2}$$

Here, the first term on the right-hand-side (i.e., $2/3$) refers to the original Hopf-like function obtained from the $T$–$\tau$ relationships in the Eddington gray atmosphere model. The second term causes the temperature to rise once the optical depth decreases below a critical level. This rise in temperature leads to an increase in the radiation pressure, which can be used to simulate magnetic pressure, and hence the presence of magnetic fields in the upper atmosphere. In addition, we introduce two physical parameters, $a$ and $b$, to specify how strong the magnetic fields are and where they are located in the atmosphere.

### 3.3. The Boundary Condition of Linear Oscillations

The mechanical surface boundary condition is a crucial factor in the propagation and reflection of stellar oscillations in the stellar atmosphere. The traditional one for the linear adiabatic oscillation, which is derived by matching to the wave solution in an isothermal atmosphere, is (W. Unno et al. 1989; J. Christensen-Dalsgaard 2008)

$$\left(1 + \frac{H_P}{r}\left[\frac{l(l+1)g}{\omega^2 r} - 4 - \omega^2 \frac{r}{g}\right]\right)\xi_r - \frac{1}{\rho g}P' = 0, \tag{3}$$

where $\xi_r$ is the radial displacement, $H_P$ is the local pressure scale height at radius $r$, $P'$ is the Eulerian perturbation of pressure, $\omega$ is the cyclic frequency of the pulsation mode, $g$ is the gravitational acceleration, and $\rho$ is density. J. Christensen-Dalsgaard (2008) proposed a reflection condition on a free boundary:

$$\delta P = P' - \rho g \xi_r = 0. \tag{4}$$

But Y. Li et al. (2021) argued that after the introduction of magnetic fields in the stellar atmosphere, the $p$-mode oscillations should be completely reflected on the magnetic-arch splicing layer. As a result, there would be no acoustic waves traveling above the magnetic-arch splicing layer. Therefore, Y. Li et al. (2021) proposed a new boundary condition on the surface of the stellar atmosphere:

$$P' = 0. \tag{5}$$

To investigate the propagation of $p$-mode stellar oscillations, we adopt the normalized kinetic energy distribution formula proposed by C. Aerts et al. (2010) and J. Christensen-Dalsgaard (2008):

$$e_k = [\xi_r^2 + l(l+1)\xi_h^2]\rho r^2, \tag{6}$$

where $\xi_h$ is the horizontal displacement.

### 3.4. Model Fittings

In the asteroseismic study of HD 49385 by S. Deheuvels & E. Michel (2011), the best-fit model was searched on every evolutionary track of stellar models within a grid of parameters, including the initial stellar mass $M$, the mixing-length parameter $\alpha$, and the initial metal abundance $Z$. In our work, we determine the initial helium abundance $Y$ by using a formula of the initial metal abundance $Z$, which is $Y = Y_0 + dY/dZ \times Z$ (A. Dotter et al. 2008). Here, we assume that the value of $Y_0 = 0.248$, and $dY/dZ = 2$ (T. Li et al. 2018). We use the formula of $[Z/X]$ to obtain the value of $Z$, which is

$$[Z/X] = \log\left(\frac{Z}{X}\right) - \log\left(\frac{Z}{X}\right)_\odot, \tag{7}$$

where we adopt the values of $(Z/X)_\odot = 0.0181$ (M. Asplund et al. 2009) in A09 models and of $(Z/X)_\odot = 0.0229$ (N. Grevesse & A. J. Sauval 1998) in GS98 models. According to S. Deheuvels & E. Michel (2011), we take the metallicity $[Z/X]$ as a fixed value of 0.09 in our subsequent calculations.

For each stellar model that meets the observational constraints, we calculate the frequency values of its oscillation modes for $l = 0$, 1, and 2. We then match the calculated frequencies with the observed ones using the method described by

$$\chi^2_{\rm all} = \frac{1}{N}\sum_{i=1}^{N}\left(\frac{\nu_i^{\rm obs} - \nu_i^{\rm mod}}{\sigma_{\nu_i^{\rm obs}}}\right)^2, \tag{8}$$

where the observed frequency is represented by $\nu_i^{\rm obs}$, the calculated frequency by $\nu_i^{\rm mod}$, and $N$ is the total number of the observed frequencies, as shown in Equation (8).

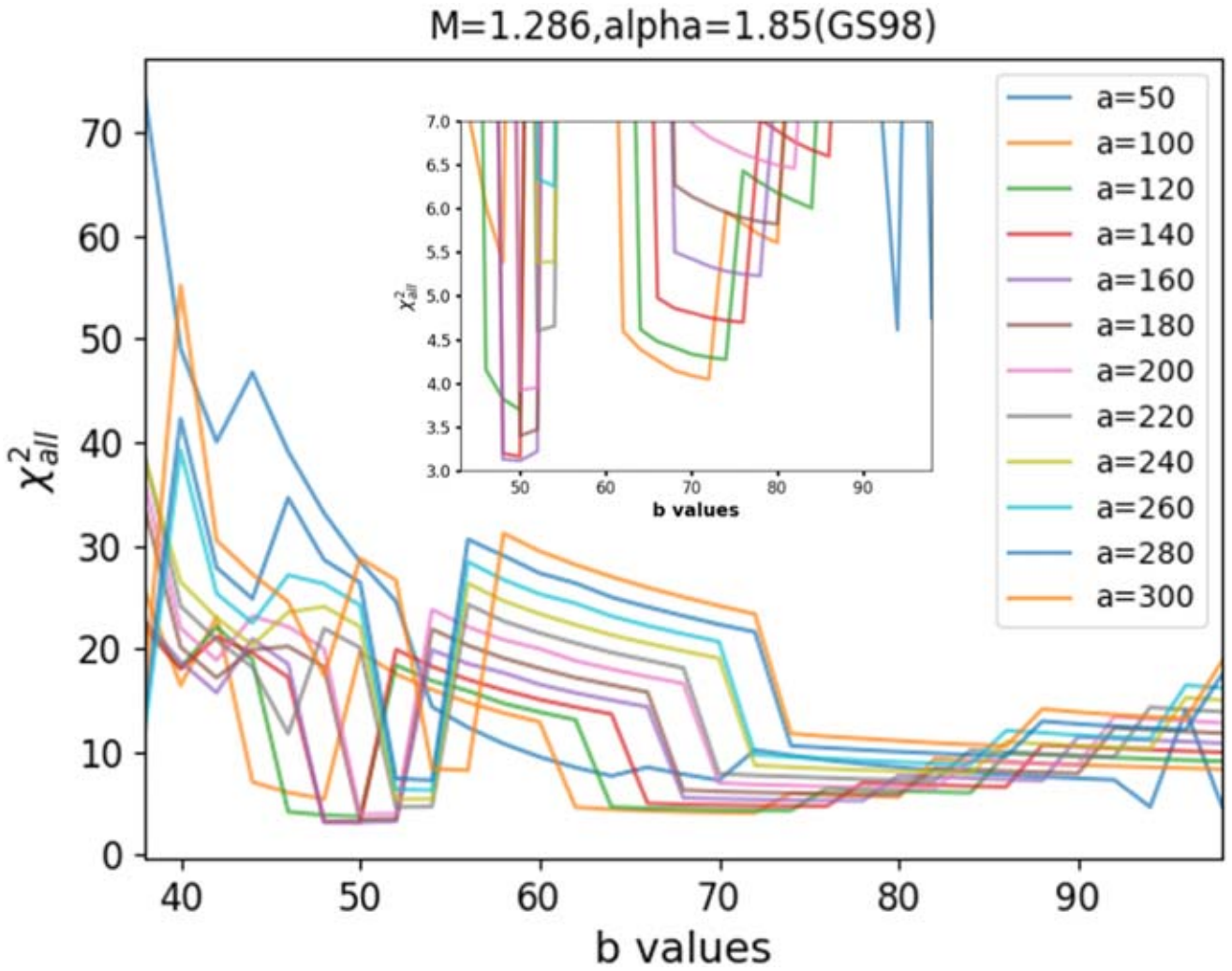

**Figure 2.** Variations of $\chi^2_{\rm all}$ with different combinations of $a$ and $b$ values in Equation (2) for the best-fitting asteroseismic models under the assumption of $M = 1.286M_\odot$ and $\alpha = 1.85$ with GS98 composition.

## 4. Fitting Results of HD 49385

When considering the extensive parameter space encompassing four parameters (i.e., $a$, $b$, $M$, and $\alpha$), it becomes evident that the computational requirements for a comprehensive grid search would far exceed the resources at our disposal. However, in our preliminary calculations, we have confirmed that the strength and location ($a$ and $b$ in Equation (2)) of the small-scale magnetic fields remain relatively unaffected by variations in the initial parameters of the stellar asteroseismic model ($M$ and $\alpha$). Consequently, we have opted for a more pragmatic approach: initially determining the values of the two free parameters ($a$ and $b$) governing the small-scale magnetic fields, and subsequently proceeding to refine our stellar models with the magnetic fields already determined in the previous step.

As an illustration, taking the chemical composition GS98 as a case in point, we embark upon the quest to find suitable initial values for $M$ and $\alpha$ as we embark on our magnetic field investigation. Initially, in the absence of magnetic field inclusion, we have conducted an extensive grid search within the following parameter range: $M = 1.231 \sim 1.342\ M_\odot$ with a step of 0.001, and $\alpha = 1.61 \sim 1.91$ with a step of 0.01. Utilizing the criterion defined by Equation (8) for the value of $\chi^2_{\rm all}$, we have established a set of fitting parameters characterized by the lowest $\chi^2_{\rm all}$ values, in which we select one combination for our initial model: $M = 1.286M_\odot$ and $\alpha = 1.85$.

Building upon this initial parameter configuration, we proceed to determine the two variables ($a$ and $b$) associated with the small-scale magnetic field. We use the parameters: $M = 1.286M_\odot$ and $\alpha = 1.85$, as the basic stellar parameters. Based on the results of Y. Li et al. (2021) for solar calculations, we scan the following grid: $a = 100 \sim 300$ with a step of 20; $b = 38 \sim 98$ with a step of 2. The results are shown in Figure 2. It can be seen that the value of $\chi^2_{\rm all}$ displays three local minima, one located at the vicinity of $a \approx 140$ and $b \approx 50$ with the least value of $\chi^2_{\rm all} \approx 3.2$, and the other two local minima with $b \approx 70$ and 90, and having a gradually larger and larger value of $\chi^2_{\rm all}$. Henceforth, we proceed with a finer grid scan with a range of $a$ from 130 to 160 with a step of 2, and of $b$ from 45 to 55 with a step size of 1. It is proved that $a = 154$ and $b = 49$ result in the least value of $\chi^2_{\rm all}$. Consequently, we have determined the parameters of magnetic fields as $a = 154$ and $b = 49$ for stellar models with GS98 chemical composition. We have performed similar calculations for stellar models with A09 chemical composition and obtained the corresponding magnetic field parameters as $a = 156$ and $b = 59$.

**Table 1**
The Calculation Grid for the Free Parameters

| Parameter | Range | Step |
|---|---|---|
| GS98 | $154\exp(-49\sqrt{\tau})$ | |
| $M$ | $1.225 \sim 1.315$ | 0.001 |
| $\alpha$ | $1.51 \sim 1.99$ | 0.01 |
| A09 | $156\exp(-59\sqrt{\tau})$ | |
| $M$ | $1.225 \sim 1.315$ | 0.001 |
| $\alpha$ | $1.51 \sim 1.99$ | 0.01 |

Under the fixed magnetic field parameters, we perform traditional asteroseismic calculations for HD 49385, with the calculation grid shown in Table 1. Table 2 shows several asteroseismic models that fit well based on Equation (8) ($\chi^2_{\rm all} < 2.6$ for stellar models with GS98 and $\chi^2_{\rm all} < 3.2$ for stellar models with A09).

The avoided crossing in HD 49385 has been extensively described in the seismological analysis by S. Deheuvels & E. Michel (2011). Based on their result, we regard the 748 $\mu$Hz oscillation mode with $l = 1$ as the avoided crossing mode and denote the observed frequency as $\nu_{\rm AC}$. Fitting this mode accurately is also the goal of our work, so we define the difference between the observed and theoretical frequencies of this mode as $D_{\nu_{\rm AC}}$. Table 2 displays the results of our models, where we choose two sets of models with the best $\nu_{\rm AC}$ fit when $\chi^2_{\rm all}$ meets the corresponding requirements for GS98 and A09 composition. Finally, we select Model E with GS98 and Model J with A09 as our best-fit models and summarize their main properties in Table 3.

## 5. Results and Discussions

Through Table 2, it becomes apparent that the GS98 models perform overall better than the A09 models on the frequency matching. When we consider the stellar observational parameters provided by S. Deheuvels et al. (2010) in Table 3, however, we find that the best-fit models with both GS98 and A09 chemical composition exhibit good agreement with the observed values of HD 49385. Compared to DM11 (S. Deheuvels & E. Michel 2011) computed with AGS05 chemical composition (M. Asplund et al. 2005) and including a core overshooting of $\alpha_{\rm ov} = 0.19$, we notice that our best-fit models have a little bit higher masses, which are therefore responsible for a shorter age due to the faster main-sequence evolution and a bit larger radius to guarantee our best-fit models having similar mean density as that of DM11. It is interesting to note that the initial metal abundance ($[Z/X] = 0.04$) for DM11 is a little lower than the observed value of 0.09 for HD 49385.

Differences between observed and theoretical frequencies are shown in Figure 3 for HD 49385. It can be seen clearly that the agreement between the frequencies of both theoretical models and the corresponding observed ones is satisfactory. Most of the oscillation modes show a frequency difference within 2 $\mu$Hz, except only for at most one mode for each value of the spherical harmonic index $l$. Compared with the results of DM11 (S. Deheuvels & E. Michel 2011), we achieve a similar but a little lower accuracy of the frequency fitting. In contrast, S. Deheuvels & E. Michel (2011) included in their frequency fitting procedure a correction of power-law type for the model frequencies to eliminate the so-called surface effect. In our work, we have not employed any type of correction for the calculated frequencies. Instead, we have included the effect of a small-scale magnetic field in the stellar photosphere.

**Table 2**
Values of the Stellar Parameters from Our Best Candidate Models

| Model | $M$ ($M_\odot$) | $\alpha$ | $T_{\rm eff}$ (K) | $\log(L/L_\odot)$ | $\log g$ | $R$ ($R_\odot$) | Age (Gyr) | $D_{\nu{\rm AC}}$ ($\mu$Hz) | $\chi^2_{\rm all}$ |
|---|---|---|---|---|---|---|---|---|---|
| | GS98 | | | | | $154\exp(-49\sqrt{\tau})$ | | | |
| A | 1.236 | 1.66 | 6039.994 | 0.649 | 3.958 | 1.931 | 4.312 | 0.652 | 2.576 |
| B | 1.234 | 1.65 | 6031.760 | 0.647 | 3.958 | 1.930 | 4.339 | 0.746 | 2.592 |
| C | 1.235 | 1.65 | 6034.327 | 0.648 | 3.958 | 1.931 | 4.324 | 0.633 | 2.593 |
| D | 1.247 | 1.70 | 6080.251 | 0.663 | 3.960 | 1.936 | 4.164 | 0.654 | 2.589 |
| E | 1.252 | 1.73 | 6101.901 | 0.670 | 3.960 | 1.938 | 4.103 | 0.566 | 2.579 |
| F | 1.253 | 1.73 | 6104.419 | 0.671 | 3.961 | 1.939 | 4.089 | 0.649 | 2.593 |
| | A09 | | | | | $156\exp(-59\sqrt{\tau})$ | | | |
| G | 1.244 | 1.58 | 5956.564 | 0.629 | 3.958 | 1.939 | 4.583 | 0.843 | 3.098 |
| H | 1.243 | 1.58 | 5953.948 | 0.628 | 3.957 | 1.938 | 4.599 | 1.229 | 3.104 |
| I | 1.245 | 1.58 | 5959.039 | 0.629 | 3.958 | 1.939 | 4.567 | 0.928 | 3.129 |
| J | 1.250 | 1.60 | 5978.259 | 0.636 | 3.958 | 1.942 | 4.497 | 0.723 | 3.134 |
| K | 1.266 | 1.67 | 6040.414 | 0.657 | 3.961 | 1.950 | 4.283 | 1.263 | 3.122 |
| L | 1.271 | 1.69 | 6058.986 | 0.664 | 3.961 | 1.951 | 4.217 | 1.002 | 3.121 |
| M | 1.239 | 1.56 | 5937.210 | 0.622 | 3.957 | 1.937 | 4.655 | 1.165 | 3.148 |

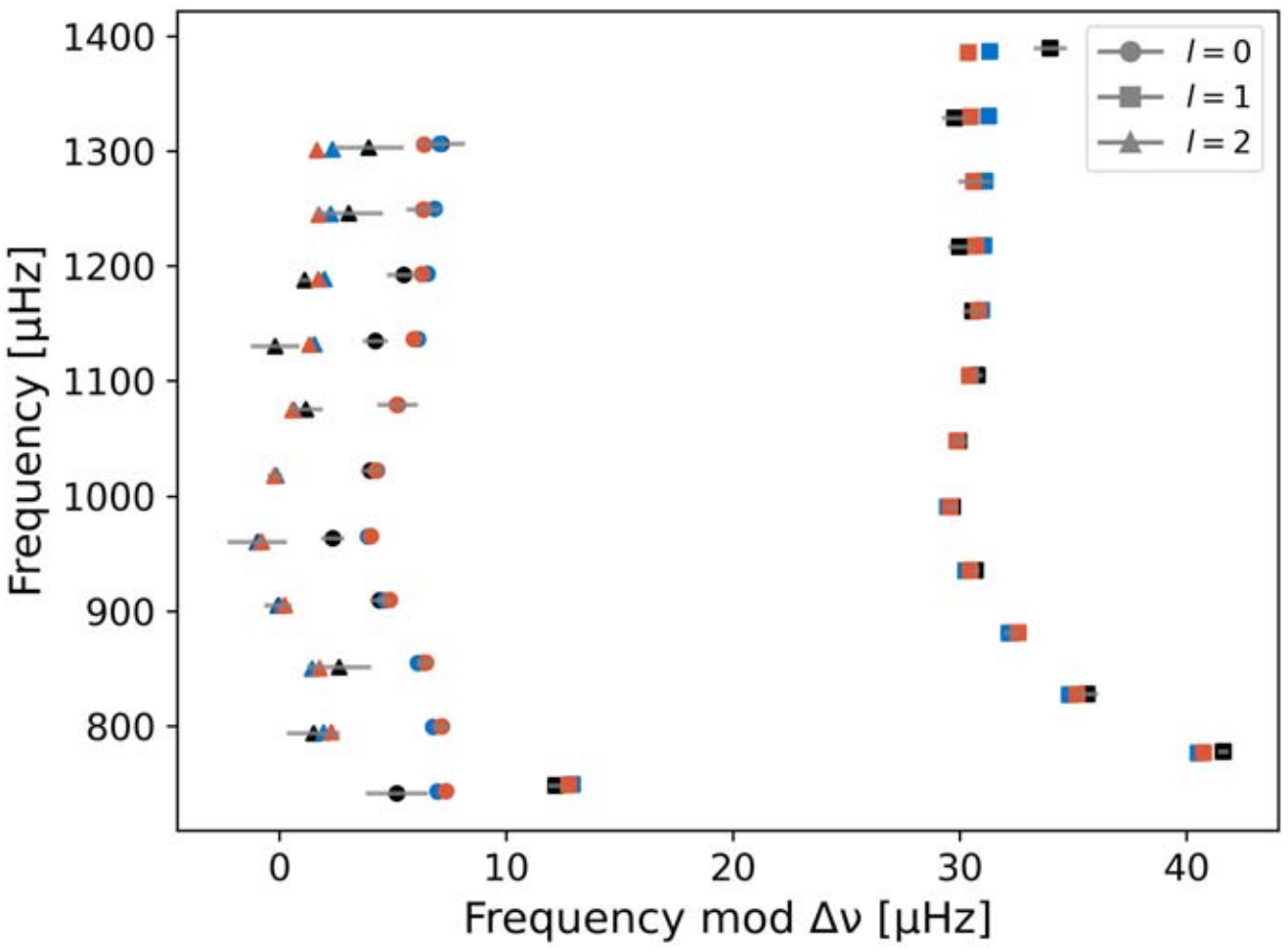

**Figure 3.** Comparisons between observed and theoretical frequencies for HD 49385. Red symbols represent the results of Model E with GS98, and blue symbols represent those of Model J with A09. The black ones with gray error bars are the observed frequencies.

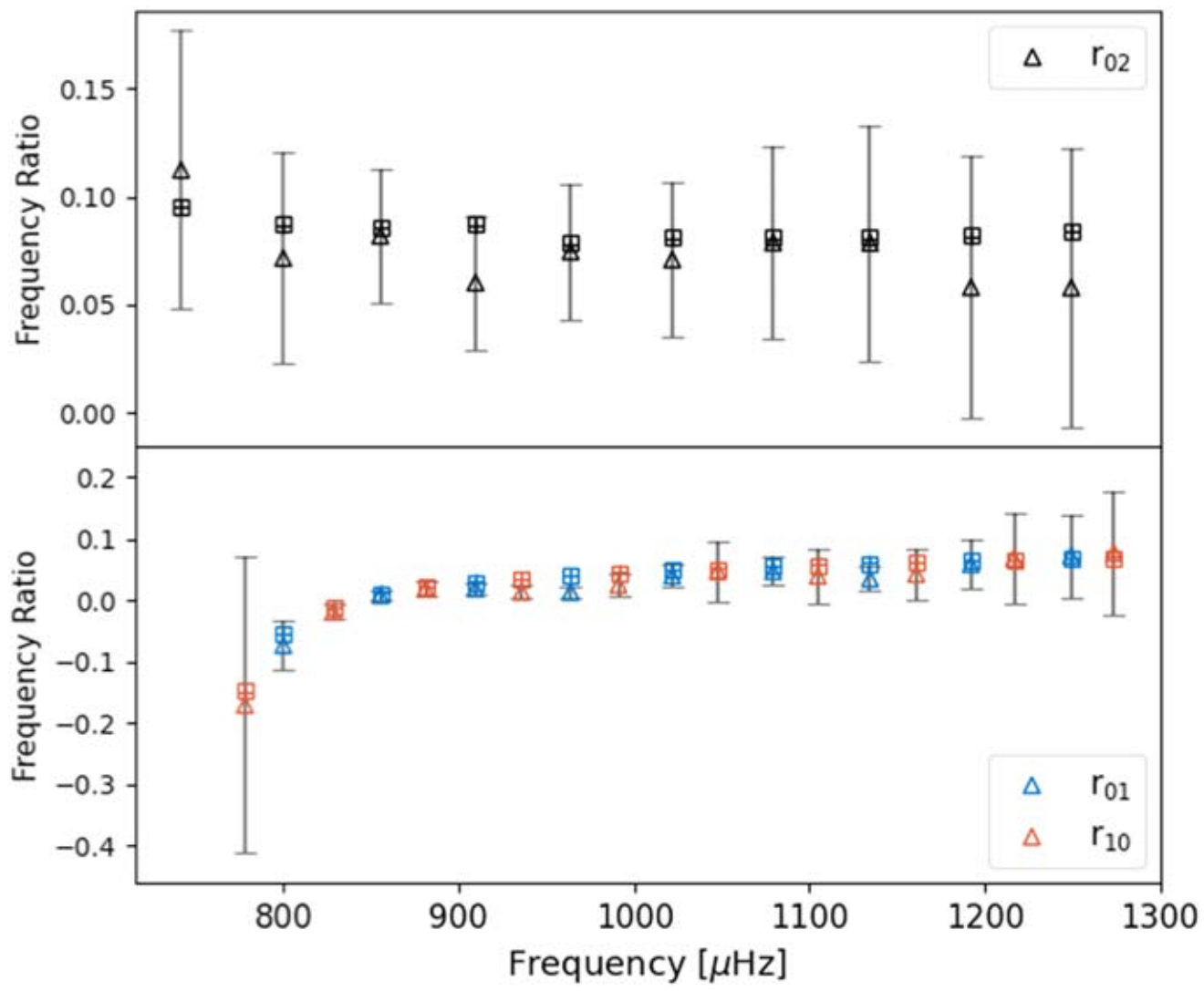

**Figure 4.** The distribution of $r_{ij}$ calculated from Equations (13)–(15). The triangles represent the observed data with error bars obtained from the error propagation formula. The squares represent the results of Model E with GS98, and the plus signs represent those of Model J with A09.

**Table 3**
The Fundamental Parameters for HD 49385

| Variables | D10 | DM11(AGS05) | GS98 | A09 |
|---|---|---|---|---|
| $T_{\rm eff}$ (K) | $6095 \pm 65$ | $66080 \pm 60$ | $6101.9^{+2.5}_{-70.1}$ | $5978.3^{+80.7}_{-41.1}$ |
| $\log g$ | $4.00 \pm 0.06$ | $3.954 \pm 0.002$ | $3.960^{+0.001}_{-0.002}$ | $3.958^{+0.003}_{-0.001}$ |
| $\log(L/L_\odot)$ | $0.67 \pm 0.05$ | ⋯ | $0.670^{+0.001}_{-0.023}$ | $0.636^{+0.028}_{-0.024}$ |
| $M$ ($M_\odot$) | ⋯ | $1.210 \pm 0.021$ | $1.252^{+0.001}_{-0.018}$ | $1.250^{+0.021}_{-0.011}$ |
| $\alpha$ | ⋯ | $0.56 \pm 0.03$ | $1.73^{+0.00}_{-0.05}$ | $1.60^{+0.09}_{-0.04}$ |
| $R$ ($R_\odot$) | ⋯ | $1.917 \pm 0.011$ | $1.938^{+0.001}_{-0.008}$ | $1.942^{+0.009}_{-0.005}$ |
| Age (Gyr) | ⋯ | $5.10 \pm 0.18$ | $4.103^{+0.236}_{-0.014}$ | $4.497^{+0.158}_{-0.280}$ |
| $M_{\rm He}$ ($M_\odot$) | ⋯ | ⋯ | 0.117 | 0.117 |
| $R_{\rm He}$ ($R_\odot$) | ⋯ | ⋯ | 0.078 | 0.078 |
| Height (km) | ⋯ | ⋯ | 1850 | 1853 |
| Strength (Gauss) | ⋯ | ⋯ | 80 | 78 |

In the work of I. W. Roxburgh & S. V. Vorontsov (2003), the importance of the large and small frequency separations ($\Delta\nu_{nl}$ and $\delta\nu$) and the ratio of small-to-large frequency separation ($r_{ij}$) is highlighted for the study of solar-like oscillations, where the large and small frequency separations

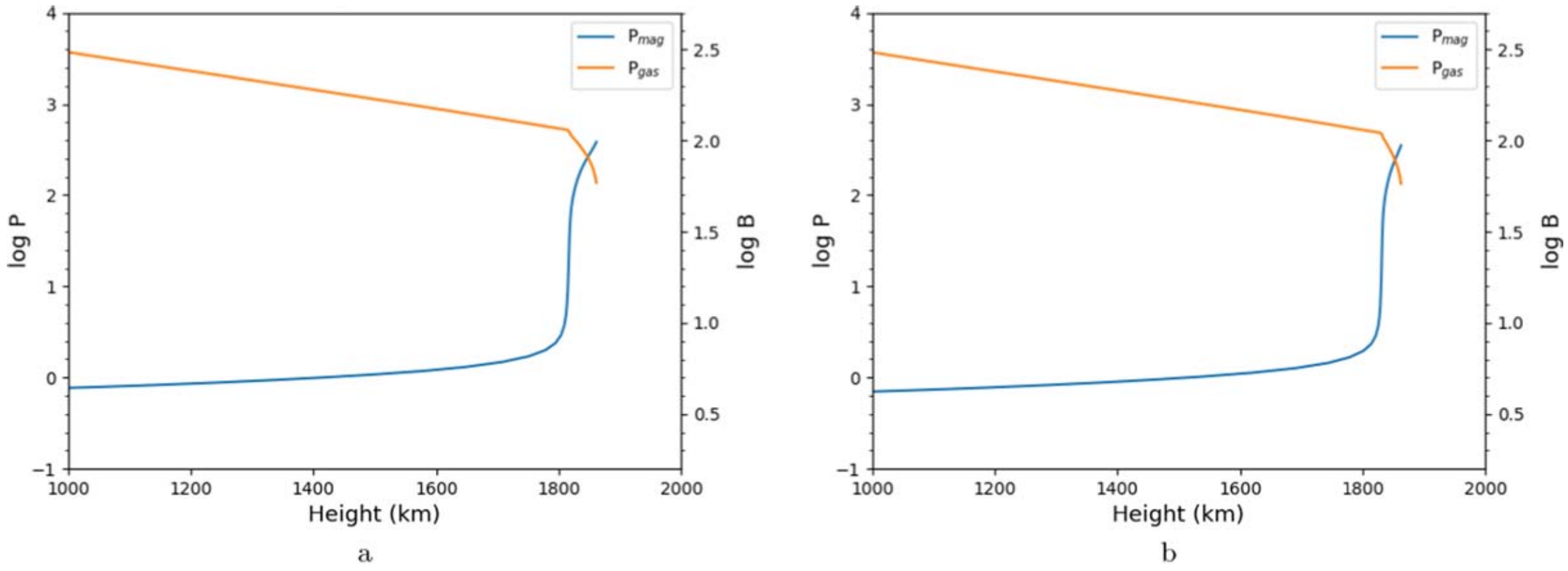

**Figure 5.** The variations of gas pressure and magnetic pressure with height. Panel (a) is for Model E with GS98, and panel (b) for Model J with A09.

are defined by

$$\Delta\nu_l(n) = \nu_{n,l} - \nu_{n,l-1}, \tag{9}$$

$$\delta\nu_{01}(n) = \frac{1}{8}(\nu_{n-1,0} - 4\nu_{n-1,1} + 6\nu_{n,0} - 4\nu_{n,1} + \nu_{n+1,0}), \tag{10}$$

$$\delta\nu_{10}(n) = -\frac{1}{8}(\nu_{n-1,1} - 4\nu_{n,0} + 6\nu_{n,1} - 4\nu_{n+1,0} + \nu_{n+1,1}), \tag{11}$$

$$\delta\nu_{02}(n) = \nu_{n,0} - \nu_{n-1,2}. \tag{12}$$

J. Christensen-Dalsgaard (1997), C. Aerts et al. (2010), and T. Wu & Y. Li (2017) demonstrated that the small frequency separations $\delta\nu_{ij}$ are determined mainly by the conditions in the stellar core. I. W. Roxburgh & S. V. Vorontsov (2003) further showed that the ratios of small-to-large frequency separations ($r_{ij}$), which is defined by

$$r_{01}(n) = \frac{\delta\nu_{01}(n)}{\Delta\nu_{l=1}(n)}, \tag{13}$$

$$r_{10}(n) = \frac{\delta\nu_{10}(n)}{\Delta\nu_{l=0}(n+1)}, \tag{14}$$

$$r_{02}(n) = \frac{\delta\nu_{02}(n)}{\Delta\nu_{l=1}(n)}, \tag{15}$$

are independent of the structure of the outer layers of the star, and can be served as a set of perfect diagnostic parameters of the stellar core for asteroseismological fitting. Figure 4 shows our results for HD 49385. It can be seen clearly that almost all of $r_{ij}$ from both best-fit models (Model E and Model J) agree well with the observed values. This result indicates that the stellar core of our best-fit models represents satisfactorily the case of HD 49385. Compared with the results of DM11 (S. Deheuvels & E. Michel 2011), which incorporate a core overshooting in their stellar models, it is still worth trying in the future to consider the overshooting effect in our stellar models for HD 49385.

C. S. Rosenthal et al. (2002) and P. S. Cally (2007) pointed out that if magnetic fields are present in the stellar atmosphere, the $p$-mode oscillations will be reflected and transmitted at an interface where the gas pressure and the magnetic pressure are approximately equal ($\beta = P_{\rm gas}/P_{\rm mag} \approx 1$). Therefore, we can see from Figure 5(a) that Model E with GS98 composition has small-scale magnetic fields of approximately 80 G and spreading concentratively at a height of about 1850 km in the photosphere. A similar analysis is also performed for Model J with A09 composition, resulting in the estimate of small-scale magnetic fields of approximately 78 G and concentrated at a height of about 1853 km as shown in Figure 5(b). It can be noticed that the two best-fit models possess almost identical properties of small-scale magnetic fields. Furthermore, it is worthwhile to compare the result of HD 49385 with the case of the Sun (Y. Li et al. 2021). The strength of the magnetic fields for HD 49385 is a little lower than that of the Sun (about 90 G), while its location is much higher than the solar case (about 630 km). The effective temperatures of the two stars are quite close (about 6000 K for HD 49385 and 5777 K for the Sun), which implies similar thermodynamic conditions and similar sizes of the convective cells in their photosphere. According to numerical simulations (see, for example, O. Steiner et al. 2008; M. Rempel 2014), the small-scale magnetic fields are closely restricted by the size of the convective cells. This would be the reason for the similar strength of the magnetic fields for the two stars. On the other hand, the radius of HD 49385 is about 1.94 times that of the Sun, which leads to a much-extended atmosphere too. This fact might account for the result that the magnetic fields of HD 49385 shift toward much higher layers in its photosphere or even chromosphere.

Finally, the most prominent feature of HD 49385 is the avoided-crossing phenomenon in its oscillation frequency spectrum. In Figure 6, we present distributions of the helium abundance, the Brunt–Väisälä frequency, and the normalized kinetic energy of three low-frequency modes with $l = 1$ for Model E as an example. As already pointed out by S. Deheuvels & E. Michel (2011), HD 49385 is an evolved star, depleting hydrogen in its core. Thereupon, it can be seen clearly in Figure 6 that the Brunt–Väisälä frequency shows a peak around the boundary of its helium core. This peak is high enough to host $g$-mode oscillations propagating in the helium core, resulting in the appearance of a mixed mode that behaves like $g$-mode in the helium core and $p$-mode in the hydrogen-abundant envelope. From Figure 6 it can be easily found that the three modes with the frequencies of 748.48, 777.91, and 828.22 $\mu$Hz indeed possess some kinetic energy in the helium core. This part of kinetic energy decreases with the increase of frequency for higher radial order modes, which is mainly due to the increase of the Brunt–Väisälä frequency near the center. On the other side, the kinetic energy quickly terminates at the boundary of the helium core for all three modes. Similar situations have been found in mixed-mode oscillations with $l = 1$ in a few pulsating red giants (X. Zhang et al. 2018, 2020), which is thought to be a good diagnostic feature of the helium core. In the present case, we estimate the helium core of HD 49385 to be 0.117 $M_\odot$ in mass and 0.078 $R_\odot$ in radius in Table 3. It can be noticed in Table 3 that the two best-fit models possess an identical helium core, although they show somewhat different features, particularly in effective temperature and radius. This fact might indicate that the frequency of the avoided-crossing mode is tightly connected with the property of the helium core since we select stellar models in Table 2 with the best $\nu_{AC}$ fit.

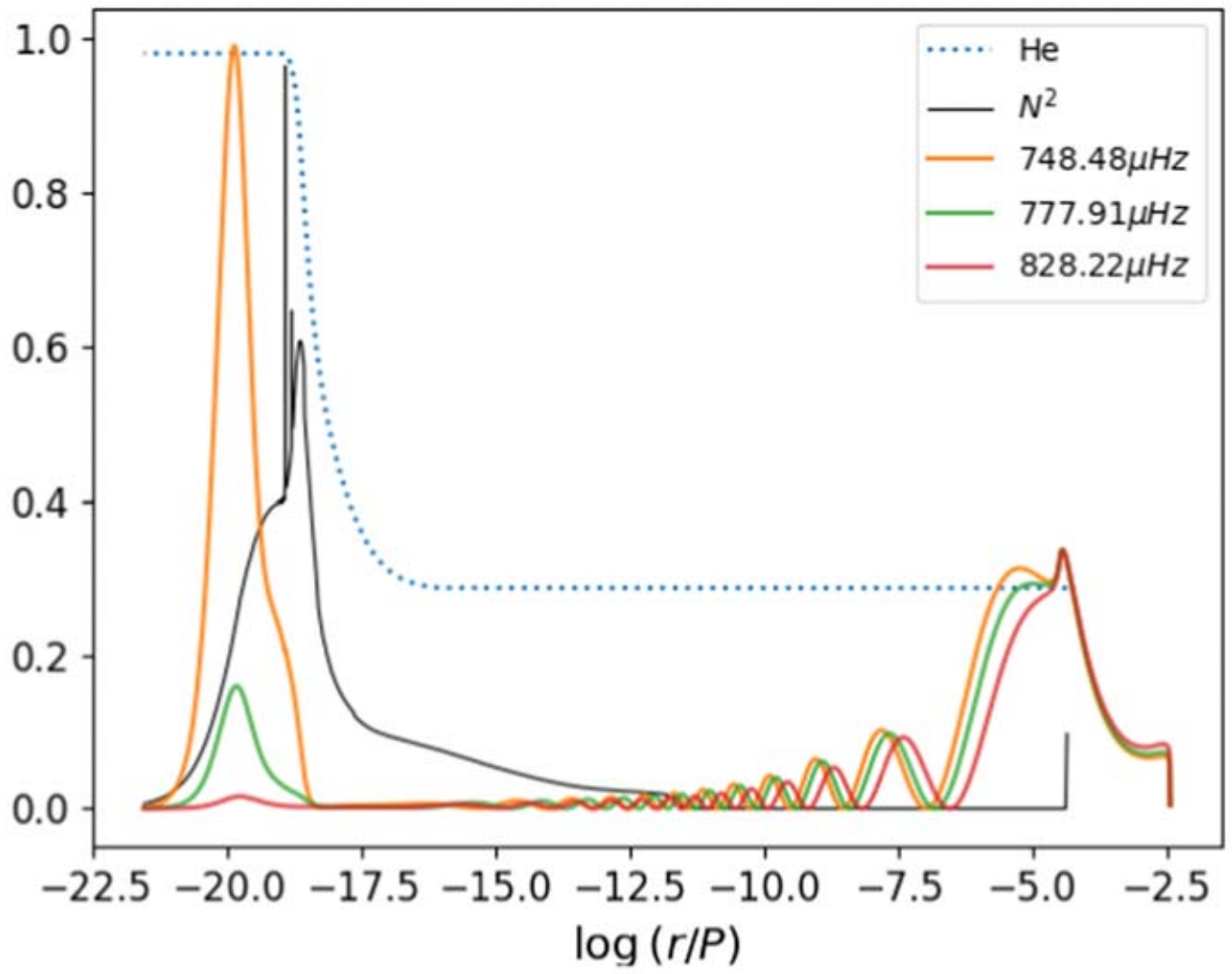

**Figure 6.** The propagation of some mixed modes in the interior of HD 49385. The black line represents the square of buoyancy frequency $N^2$, the blue dotted line corresponds to the distribution of He abundance, and the remaining three solid lines represent the normalized kinetic energy for $l = 1$ modes with frequencies of 748.48, 777.91, and 828.22 $\mu$Hz.

## 6. Conclusions

HD 49385 was observed by the space telescope CoRoT, resulting in well-resolved oscillation modes with $l = 0$, 1, and 2. A very interesting phenomenon, namely the avoided crossing, was found for $l = 1$ modes, which was investigated extensively by S. Deheuvels & E. Michel (2011). However, all of the best-fit theoretical models currently available are unable to match all the asteroseismic signatures observed on HD 49385. The small-scale magnetic fields are a crucial component in the stellar atmosphere. However, they have not been considered in traditional asteroseismic models. Inspired by the work of Y. Li et al. (2021) on the small-scale magnetic fields of the Sun, we have incorporated a modified Eddington $T$–$\tau$ relation to mimic the effect of the small-scale magnetic fields in the asteroseismic models for HD 49385, and the main conclusions are summarized below.

1. The best-fit models with the small-scale magnetic fields show a much better fit for the observed frequencies than those without the small-scale magnetic fields. More accurate stellar parameters, such as $T_{\rm eff}$, $\log g$, mass, radius, and luminosity, have been obtained, and are in good agreement with the observed values.
2. Our best-fit models with both GS98 and A09 chemical composition indicate the presence of small-scale magnetic fields in the atmosphere of HD 49385. The small-scale magnetic fields concentrate at a height of approximately 1850 km and possess a mean strength of about 80 G. Compared with the case of the Sun, the small-scale magnetic fields at present are a little weaker but located much higher, consistent with the fact that HD 49385 is an evolved star with a similar effective temperature but a much larger radius compared with the Sun. The present study can also serve as a practical method to estimate the properties of the small-scale magnetic fields on remote stars, which cannot be directly measured by traditional spectroscopic or photometric methods.
3. With the aid of selecting the best-fit models well fit to the avoided-crossing mode 748.48 $\mu$Hz, we have confirmed that the frequency of the avoided-crossing mode is tightly connected with the property of the helium core. Our best-fit models with both GS98 and A09 chemical composition show identical mass and radius of the helium core, resulting accurate estimate of the corresponding values: 0.117 $M_\odot$ in mass and 0.078 $R_\odot$ in radius. These properties restrict accurately the age of HD 49385 to be 4.1 Gyr for GS98 composition and 4.5 Gyr for A09 composition.

## Acknowledgments

This work has been cosponsored by National Natural Science Foundation of China (grant Nos. 12288102 and 12133011), National Key R&D Program of China (grant No. 2021YFA1600400/2021YFA1600402), and the International Center of Supernovae at the Yunnan Key Laboratory (grant No. 202302AN360001). T.W. thanks for financial support from the B-type Strategic Priority Program of Chinese Academy of Sciences (grant No. XDB41000000), National Natural Science Foundation of China (grant No. 12273104), Yunnan Fundamental Research Projects (grant No. 202401AS070045), Youth Innovation Promotion Association of Chinese Academy of Sciences, and Yunnan Ten Thousand Talents Plan Young & Elite Talents Project.

*Software:* MESA (v11554; B. Paxton et al. 2011, 2013, 2015, 2018, 2019).

## ORCID iDs

Yaguang Li https://orcid.org/0000-0003-3020-4437
Yan Li https://orcid.org/0000-0002-1424-3164
Guifang Lin https://orcid.org/0000-0001-6725-8207
Tao Wu https://orcid.org/0000-0001-6832-4325